\documentstyle[pre,aps]{revtex}

\begin{document}
\draft
\wideabs{
\title{Dielectric relaxation in chevron surface stabilized ferroelectric 
liquid crystals}
\author{W. Je\.zewski, W. Kuczy\'nski, and J. Hoffmann}
\address{Institute of Molecular Physics, Polish Academy of Sciences,\\ 
Smoluchowskiego 17, 60-179 Pozna\'n, Poland}
\date{9 November 2005}
\maketitle

\begin{abstract}
The dielectric response of surface stabilized ferroelectric liquid crystals 
with chevron layer structure is studied within low and intermediate frequency 
ranges, characteristic for collective molecular excitations. By analytically 
solving the dynamic equation for collective molecular fluctuations under 
a weak alternating electric field, it is demonstrated that chevron cells 
stabilized by both nonpolar and polar surface interactions undergo at medium 
frequencies two Debye relaxation processes, connected with two chevron slabs, 
on opposite sides of the interface plane. This result is confirmed, 
experimentally, making use of the electro-optic technique. Based on 
qualitative arguments supported by microscopic observations of zigzag defects 
at different frequencies and amplitudes of the external electric field, it is 
shown that, at low frequencies, the electro-optic response of chevron samples 
is determined by three kinds of motions of zigzag walls. The first two 
dynamic categories are related to collective relaxation processes at weak 
fields, within smectic A layers forming zigzag walls, and drift or creep 
motions of thick walls occuring at stronger field amplitudes. Dynamic 
processes of the third kind correspond to sliding of zigzag walls, which 
appear at yet stronger field amplitudes, but below the switching threshold.  
 
\end{abstract}

}

\section{Introduction}
Thin ferroelectric liquid crystal cells with bookshelf and chevron structures 
play the key role in developing high-resolution and large-area display screen 
technologies\cite{la}. The main technical advantage of these 
systems is the possibility of high-speed electro-optic switching between 
bistable orientational states stabilized by surface 
interactions\cite{c1,h1,h2,ri,xu,co}. Consequently, studies of the dynamic 
behavior of surface stabilized ferroelectric liquid crystals (SSFLC) have 
mostly been focused on ferroelectric switching processes. In the case of 
the chevron geometry, a challenging problem is the inclusion of the process of 
a rapid reorientation of molecules at the chevron interface in the theoretical 
description of switching phenomena\cite{m1,m2,br,ha}. As compared to 
investigations of the switching dynamics, the behavior of chevron 
systems in the presence of the alternating external electric field of small 
amplitudes, less than the switching threshold, has not been intensively studied. 
Collective molecular excitations generated by a weak sinusoidal applied voltage 
have been analyzed theoretically by solving a one-dimensional field equation 
describing the dynamic distribution of the azimuthal angle of the molecular 
$c$ director (the projection of the molecular director on the smectic plane) 
with respect to the axis perpendicular to boundary plates, and then by 
determining the dielectric permittivity\cite{p1,ka}. However, the solution to 
this equation has been derived by assuming an unrealistic condition that, at the 
chevron interface, the azimuthal angle of the $c$ director does not depend on 
the external field\cite{p1,ka}. Clearly, such an assumption is equivalent to 
the requirement that the strength of the interface anchoring interactions is 
infinite.

In this paper, the dynamic equation that describes collective molecular 
excitations in uniform symmetric SSFLC with the chevron structure under the 
influence of a weak sinusoidal electric field is solved for more natural 
interface anchoring conditions, allowing the azimuthal orientation of 
molecules to depend on the applied electric field also at the chevron 
interface. The surface interactions are assumed to include both nonpolar and 
polar couplings. In general, the existence of polar surface interactions, in 
addition to the nonpolar ones, leads to a difference in anchoring energies on 
two bounding plates. It is proved that the difference between the surface 
energies can be modified by applying the electric field across boundary 
plates. As will be shown, the alternating voltage induces in uniform SSFLC 
with bookshelf and chevron structures a nonuniform contribution to the long 
time average of the azimuthal angle. Such a change in the time average of 
molecular orientation gives rise to  a variation of anchoring energies on cell 
plates, compared to the case of the stationary molecular orientation. It is 
argued here that the differentiation of effective surface energies, changed by 
a weak alternating voltage, causes collective relaxation processes in two 
slabs of uniform chevron SSFLC to run in distinct manners, with separate 
relaxation times. By imposing a constraint on the effective anchoring 
energies, these relaxation times can be interrelated. Resulting calculated 
dielectric loss spectra are shown to be in qualitative agreement with 
experimental spectra, obtained by employing the electro-optic 
method\cite{pi,ku}.

Experimental investigations of chevron SSFLC in the low-frequency regime have 
revealed a strong dependence of the electro-optic response on the voltage 
amplitude, even for small amplitudes, much less than the switching 
threshold\cite{p2}. By carrying out microscopic observations and by performing 
measurements of the electro-optic loss, it is shown here that the low-frequency 
electro-optic response of chevron samples is determined by the dynamics of 
zigzag walls\cite{ri,c2,li}. Dynamical processes occuring at very weak fields 
are identified with collective relaxation motions of molecules in smectic A 
layers, which form thin and thick zigzag walls of various sizes. It is found 
that, in the low-frequency regime, these motions can be described by a spectrum of 
Debye processes characterized by a continuous, powerlike distribution of 
relaxation times. The change of the low-frequency electro-optic loss as the 
voltage increases is attributed to the appearance of creeping motions of thick 
zigzag walls. Dynamic processes responsible for changes of the low-frequency 
electro-optic loss under further growth of the voltage are ascribed to sliding 
motions and gradual disappearance of zig-zag walls.
\section{Theoretical analysis of the dielectric response at intermediate 
frequencies}
\subsection{Director equation of motion}
Cooperative dynamics in bookshelf and chevron cells is usually studied by 
analyzing fluctuations of the azimuthal angle $\phi$ in the smectic plane, 
along the surface normal direction $X$\cite{h2,ri,p1}, as illustrated in 
Fig. \ref{f1}. The contribution to the dielectric response, due to 
fluctuations of the angle $\phi$ under an applied alternating field, is 
usually significant in a wide range of medium frequencies. Therefore, this 
range is called here the intermediate frequency region. Assuming appropriate 
boundary conditions for the azimuthal angle, its fluctuations can be described 
independently in each chevron slab\cite{h2,ri}, by means of a dynamic field 
equation. In the presence of the sinusoidal external electric field applied 
along the $X$ axis and alternating with the frequency $\omega$, this equation 
reads  
\begin{equation}
\frac{\partial^2\phi}{\partial x^2}-\gamma\frac{\partial\phi}{\partial t}=
A\sin\phi\cos\omega t\,\,\,,
\end{equation}
where $\gamma=\gamma_\phi/K$ with $\gamma_\phi$ being the azimuthal viscosity 
and $K$ being the twist elastic constant. The quantity $A=P_sE/K$ with $P_s$ 
denoting the local spontaneous polarization and $E$ being the electric field 
amplitude. Boundary conditions for Eq. (1) should involve preferred 
(zero-field) angle values at border surfaces as well as at the chevron 
interface. In uniform chevron cells, the $c$ director is pretilted 
(in the absence of the external field) by $\pm\phi_0$, where the convention 
that $+$ and $-$ refer respectively to quantities on upper and lower chevron 
slabs is used (this convention will henceforth be adopted). Consequently, for 
the zero-field equilibrium state of uniform chevrons, the azimuthal angle $\phi$ 
is required here to obey the customary relation
\begin{equation}
\phi_\pm(0)=\mp\phi_0\,\,\,,
\end{equation}
where $\phi_\pm (0)$ refers to values of $\phi$ immediately above and below 
the interface, respectively. The amount of the pretilt $\phi_0$ is, in general, 
nonzero and depends on the molecular and layer tilt angles and surface 
interaction strengths\cite{m1}. For the downward orientation of the 
polarization (as shown in Fig. \ref{f1}), the anchoring surface energies can 
be supposed in the form\cite{h1,h2,m1,na} 
\begin{equation}
W_\pm(\phi)=-\gamma_1\cos^2(\phi\pm\phi_0)\pm\gamma_2\cos(\phi\pm\phi_0)\,\,\,,
\end{equation}
where $\gamma_1$ and $\gamma_2$ are respectively the nonpolar and polar 
surface energy strengths, the same on both bounding plates. Hence, the boundary 
conditions for $\phi$ near its zero-field equilibrium value $\phi_0$ can be 
expressed in the case of downward-oriented chevron cells as\cite{h1,m1,m2}
\begin{equation}
\frac{\partial}{\partial x}\phi_{\pm}(x=\pm\frac{d}{2})\simeq\pm v_\pm
[\phi_\pm(x=\pm\frac{d}{2})\pm\phi_0]
\end{equation}
with $d$ denoting the thickness of a chevron cell and 
\begin{equation}
v_\pm=2\lambda_1\mp\lambda_2
\end{equation}
being the combined surface interactions, where $\lambda_1=\gamma_1/K$ and 
$\lambda_2=\gamma_2/K$. As the relations (4) and (5) show, the spatial 
derivative of $\phi$ depends on the combined surface interactions (taking for 
$\gamma_2\ne 0$ different values at top and bottom boundary plates). For weak 
external fields and strong anchoring interactions at boundary and interface 
surfaces, one can assume, in addition to the condition (4), that\cite{m1,m2}
\begin{equation}
\phi_\pm(\pm d/2)=\phi_\pm (0)\,\,\,.
\end{equation}

In the static case ($\omega=0$), Eq. (1) reduces to the Euler-Lagrange equation
\begin{equation}
\frac{\partial^2}{\partial x^2}\phi=A\sin\phi\,\,\,.
\end{equation}
Using conditions (4) and (6), one finds in the case of downward oriented 
chevron cells in a weak constant field the following solution of Eq. (4)
\begin{equation}
\phi_\pm(x)=\mp\phi_0+c f_\pm(x)+{\cal O}(A^2)\,\,\,,
\end{equation}
where 
\begin{equation}
f_\pm(x)=\pm\frac{A}{2}\sin\phi_0(\frac{d}{2v_\pm}\pm\frac{d}{2}x-x^2)\,\,\,.
\end{equation}
According to Eqs. (5) and (9), the field-induced distribution of $\phi(x)$, 
determined by the functions $f_\pm$, is different in the top and bottom 
chevron slabs, provided that $A\ne 0$ (in particular, 
$|\phi_+(0)|\ne|\phi_-(0)|$). This is a result of the difference between 
$v_+$ and $v_-$.

For weak alternating electric fields, the solution of Eq. (1), in the case of 
downward polarization orientation, can be written as
\begin{eqnarray}
\phi_\pm(x,t)=&\mp&\phi_0+b_0^\pm+b_1^\pm x\nonumber\\
&+& F_\pm(x)\cos(\omega t-\beta_\pm)+{\cal O}(A^2)\,\,\,,
\end{eqnarray}
with $F_\pm(x)$ having series expansions
\begin{equation}
F_\pm(x)=\sum_{l=0}^{\infty}a_l^\pm x^l\,\,\,,
\end{equation}
where $\beta_\pm$ represents field phase shifts, different, in general, 
in the top and the bottom chevron slabs, $b$'s and $a$'s are spatially and 
temporary independent coefficients. Both $b_0^\pm$ and $b_1^\pm$ do not 
depend also on $A$, while $a_l^\pm\sim A$, $l=0,1,2,...\,$. It is easy to verify 
that Eq. (1) has no solution of the form (10) for $b_1^\pm=0$ and for any 
nonzero $b_1^\pm$ such that $b_1^\pm\rightarrow 0$ as $A\rightarrow 0$.
Inserting (10) into (1), equating components with the same time functions and 
with the spatial variable increased to the same powers, one obtains an 
infinite sequence of relations
\parbox{7.5cm}{\begin{eqnarray*}
(l+2)!\,a_{l+2}^\pm\cos\beta_\pm-l!\,\gamma\omega a_l^\pm\sin\beta_\pm&=&A\rho_l^\pm 
(b_1^\pm)^{l-2}\,\,\,,\\
(l+2)!\,a_{l+2}^\pm\sin\beta_\pm+l!\,\gamma\omega a_l^\pm\cos\beta_\pm&=&0\,\,\,,
\end{eqnarray*}} \hfill 
\parbox{1cm}{\begin{eqnarray}\end{eqnarray}}\\
where $l=0,1,2,...,$ and 
\begin{equation}
\rho_l^\pm=
\left\{
\begin{array}{cr}
 (-1)^{(l+2)/2}\sin(\mp\phi_0+b_0^\pm)\,\,\,,\mbox{if}\,\,\, 
 l=0,2,4,...\,\,\,,\nonumber\\
 \\
 (-1)^{(l+1)/2}\cos(\mp\phi_0+b_0^\pm)\,\,\,,\mbox{if}\,\,\, 
 l=1,3,5,...\,\,\,.\nonumber
\end{array}
\right. 
\end{equation}
Solving Eqs. (12) with respect to the coefficients $a$ yields
\begin{equation}
a_l^\pm=\frac{1}{l!}A\rho_l^\pm\cos\beta_\pm(b_1^\pm)^{l-2}\,\,\,,\,\,\,
l=0,1,2,...,
\end{equation}
with
\begin{equation}
(b_1^\pm)^2=\frac{\gamma\omega}{\tan\beta_\pm}\,\,\,. 
\end{equation}
Thus, according to (13) and (14), the functions $F_\pm$ can be written in a 
concise form as 
\begin{equation}
F_\pm(x)=-A(b_1^\pm)^{-2}\cos\beta_\pm\sin(\mp\phi_0+b_0^\pm+b_1^\pm x)\,\,\,.
\end{equation}
The simple form of $F_\pm(x)$ is a result of the occurrence of field-independent 
components (being a solution of the Laplace equation 
$\frac{\partial^2}{\partial x^2}\phi=0$) in $\phi_\pm(x,t)$. The existence of 
these linear components in (10) implies that the long time averages 
$\bar\phi_\pm(x)=\lim_{T\rightarrow\infty}T^{-1}\int_{0}^{T}\phi_\pm(x,t)dt$ 
exhibit nonuniform distributions, given by 
$\bar\phi_\pm(x)=\mp\phi_0+b_0^\pm+b_1^\pm x$. Such a dynamic splay is a 
consequence of heterogeneous, throughout chevron cells, reaction of viscosity 
interactions to the action of alternating electric field (note that 
$b_1^\pm\rightarrow 0$, and thereby $b_0^\pm\rightarrow 0$, as 
$\gamma\rightarrow 0$). Hence, average values of $\phi_\pm$ at boundary and 
interface surfaces depend not only on boundary and interface anchoring 
couplings but also on viscosity and elastic interactions (via $\gamma$). 

Since the dynamic splay is generated by the applied alternating voltage, one 
can assume that, for small voltage amplitudes, this splay is not significant 
and hence high-order components (including coefficients $a_l^\pm$ with large 
powers of $b_1^\pm$) in expansions (11) can be neglected. Moreover, the 
functions $F_\pm$ should approach their static counterparts $f_\pm$ as the 
field frequency decreases. Consequently, $F_\pm$ can be approximated as
\begin{equation}
F_\pm(x)\simeq a_0^\pm+a_1^\pm x+a_2^\pm x^2\,\,\,.
\end{equation}
By means of relation(3), the boundary conditions for solution (10) are 
given now by\cite{m1,m2}
\begin{equation}
\frac{\partial}{\partial x}\phi_{\pm}(x=\pm\frac{d}{2},t)\simeq\mp\tilde{u}_\pm\mp 
u_\pm F_\pm(\pm\frac{d}{2})\cos(\omega t-\beta_\pm)\,\,\,,
\end{equation}
where  
\begin{eqnarray}
\tilde{u}_\pm&=&\lambda_1\sin(2b_0^\pm\pm b_1^\pm d)\mp\lambda_2\sin(b_0^\pm\pm 
b_1^\pm\frac{d}{2})\,\,\,,\\
u_\pm&=&2\lambda_1\cos(2b_0^\pm\pm b_1^\pm d)\mp\lambda_2\cos(b_0^\pm\pm 
b_1^\pm\frac{d}{2})\,\,\,.
\end{eqnarray}
Then, imposing these boundary conditions on $\phi_{\pm}(x,t)$ and by analogy 
with the condition (6), using the constraint that $F_\pm(\pm d/2)=F_\pm(0)$ 
gives
\begin{eqnarray}
&&b_1^\pm=\mp\tilde{u}_\pm\,\,\,,\\
&&(b_1^\pm)^2=\frac{4u_\pm}{d}\,\,\,,\\
&&\tan(\mp\phi_0+b_0^\pm)=\pm\frac{4}{b_1^\pm d}\,\,\,.
\end{eqnarray}
Relations (22), or alternatively (21), and (23) enable one to express the 
constants $b_0^\pm$ and $b_1^\pm$ by the material parameters $\lambda_1$, 
$\lambda_2$, and $\phi_0$. Using, in turn, Eqs. (14), (22), and (23), one 
finds  
\begin{equation}
F_\pm(x)\simeq-\frac{A}{2}\sin(\mp\phi_0+b_0^\pm)\cos\beta_\pm
(\frac{d}{2u_\pm}\pm\frac{d}{2}x-x^2)\,\,\,.
\end{equation}     
The distribution of the azimuthal angle given by the above relation is 
similar as in the static case [Eq. (9)], but involves modified, due to the 
action of the alternating electric field, {\it effective} combined surface 
interactions $u_\pm$, instead of $v_\pm$.   

In the case of strong surface anchoring, one can assume that, at boundary 
surfaces, the effect of the dynamic splay is roughly insignificant. Then, 
$\bar{\phi}_\pm(\pm d/2)\simeq\mp\phi_0$, and hence,
\begin{equation}
b_0^\pm\simeq\mp b_1^\pm\frac{d}{2}\,\,\,.
\end{equation}
Thus, the combined surface interactions $u_\pm$ are close to $v_\pm$, 
respectively, for systems in weak applied fields and with strong anchoring of 
molecules at surfaces. The relation (25), together with Eqs. (22) and (23), 
leads to an approximate relationship between $\lambda_1$, $\lambda_2$, and 
$\phi_0$. Unfortunately, this relationship is not uniquely determined, since 
Eqs. (23) and (25) have many solutions for a given $\phi_0$. However, instead 
of assuming the constraint $F_\pm(\pm d/2)=F_\pm(0)$, which refers to the 
static case rather than to the dynamic case, one can use the condition that 
the long time average of the interface energy takes its minimum value. This 
condition concerns the dependence of the average interface energy on the 
average azimuthal angle $\bar{\phi}_\pm(0)$. According to (25), the average 
azimuthal angle takes at chevron interface the values
\begin{equation}
\bar{\phi}_\pm(0)\simeq\pm\phi_0\mp b_1^\pm\frac{d}{2}\,\,\,.
\end{equation}
Since, in general, $b_1^+\ne b_1^-$, the abrupt change of $\phi$ occurring for 
$E=0$ at the interface is on average modified when the dynamic splay appears 
(in the presence of applied alternating field). Such a modification of the 
discontinuity of the average azimuthal angle at the interface is connected 
with an increase of the average interface energy. For chevron structures with 
small layer and molecular tilt angles (i.e., for typical chevron cells), the 
growth of the density of this energy, $W_I$, can be determined by
\begin{equation}
\varepsilon\varepsilon_0W_I=P_s^2-P_x^+P_x^-+P_y^+P_y^-\,\,\,,
\end{equation}
where $\varepsilon$ is the dielectric permittivity, whereas $P_x^\pm$ and 
$P_y^\pm$ denote $x$ and $y$ components of the spontaneous polarization of 
molecules immediately above and below the interface plane, respectively. 
Assuming that $|b_0^\pm|$ are small compared with $|\phi_0|$, allows $W_I$ 
to approximate as follows
\begin{equation}
\varepsilon\varepsilon_0W_I\simeq\frac{1}{2}(b_0^++b_0^-)^2-\frac{1}{4}
(b_0^+b_0^-)^2\,\,\,,
\end{equation}
By minimizing $W_I$ with respect to $b_0^-$ (for fixed $b_0^+$) one gets
\begin{equation}
b_0^-\simeq-\frac{b_0^+}{1-\frac{1}{2}(b_0^+)^2}\,\,\,.
\end{equation}
Consequently, by virtue of (22) and (25), the above constraint yields
\begin{equation}
u_-\simeq\frac{u_+}{1-du_+}\,\,\,.
\end{equation}
This approximate relation is valid for systems with strong, asymmetric 
anchoring of molecules at bounding plates, i.e., for large 
$\gamma_1$ and $\gamma_2$ (in comparison to $K$), such that 
$2\gamma_1\ne\gamma_2$. As it is shown below, it is very useful for 
determining the dielectric response of chevron samples.
\subsection{Dielectric response}
To determine the dynamic dielectric susceptibility and thereby the dynamic 
dielectric response of chevron SSFLC, the solution (10) to Eq. (1) is applied. 
Thus, in the case of a sinusoidal external electric field $E(t)$ acting along 
the $X$ axis, the azimuthal mode contribution to the dynamic dielectric 
susceptibility can be written as 
\begin{eqnarray}
\chi(\omega)= 
&&\lim_{E\rightarrow 0}\frac{P_s}{E(t)}{\big [}{\big <}
\cos\phi_+(x,t){\big >}_+-{\big <}\cos\bar{\phi}_+(x){\big >}_+\nonumber\\
&&+{\big <}\cos\phi_-(x,t){\big >}_--{\big <}\cos\bar{\phi}_-(x){\big >}_-
{\big ]}\,\,\,,
\end{eqnarray}
where $<\!\!...\!\!>_\pm$ denote space averages over respectively upper and lower 
chevron arms, so that, e.g.,
\begin{equation}
{\big <}\cos\phi_\pm(x,t){\big >}_\pm=\pm\frac{2}{d}\int^{\pm d/2}_{0}\!\!\!
\cos\phi_\pm(x,t)dx\,\,\,.
\end{equation}
Using Eqs. (10), (22)-(25), and (31), and assuming that $|b_0^\pm|$ are small, 
one derives
\begin{equation}
\frac{\chi(\omega)}{\varepsilon_0}=\varepsilon'(\omega)+
\varepsilon''(\omega)\tan(\omega t)\,\,\,,
\end{equation}
with the dielectric permittivity
\begin{equation}
\varepsilon'(\omega)=B(\cos^2\beta_++\cos^2\beta_-)
\end{equation}
and the dielectric loss permittivity
\begin{equation}
\varepsilon''(\omega)=B(\tan\beta_+\cos^2\beta_++\tan\beta_-\cos^2\beta_-)\,\,\,,
\end{equation}
where
\begin{equation}
B=\frac{Ad}{4u_+}\sin^2\phi_0\,\,\,.
\end{equation}
By means of Eqs. (15) and (22), the phase shifts $\beta_\pm=\omega\tau_\pm$ 
with the relaxation times
\begin{equation}
\tau_\pm=\frac{\gamma d}{4u_\pm}\,\,\,.
\end{equation}
Hence, the dielectric response of chevron cells stabilized by both nonpolar 
and polar surface interactions can be described in the case of weak applied 
electric field by two Debye processes. These processes are determined by 
different relaxation times, each characterizing collective motions of 
molecules on opposite sides of the chevron interface. Making use of 
relation (30), valid for systems that display strong anchoring of molecules 
at bounding surfaces, one can describe the dielectric response of chevron 
samples by three parameters: $B$, $\tau_+$ (or equivalently $\tau_-$), and 
$d$. Moreover, having determined the relaxation times, one obtains the 
parameters $\gamma$, $u_+$, and $u_-$. A simple method to calculate the 
combined surface interactions $u_+$ and $u_-$, based on experimental data, 
is described below.   
\subsection{Calculation of surface energy parameters}
The permittivity spectra (34) and (35) involve two Debye relaxation processes 
with equal strengths. In such a case, the loss permittivity possesses a single 
maximum if the ratio $\nu=\tau_+/\tau_-$ is $\nu\ge\nu_c$, where 
$\nu_c=3-2\sqrt{2}\approx 0.172$, and two maxima of equal heights if 
$\nu<\nu_c$. However, the case of the double maximum form of 
$\varepsilon''(\omega)$ is not considered here because typical experimental loss 
permittivity spectra obtained for chevron cells do not display equal, or nearly 
equal, maxima that could be ascribed to collective azimuthal excitations of 
molecules.

In order to compare theoretical [Eqs. (34) and (35)] and experimental spectra 
$\varepsilon(\omega)$, one has to determine combined surface interactions 
$u_\pm$. By applying the approximate relation (30), one can describe 
$\varepsilon(\omega)$ by one energy parameter, say $u_+$. Then, the position 
of the maximum of $\varepsilon''(\omega)$ is given by
\begin{equation}
\omega_{\rm max}\tau_+=1/\sqrt{\nu}\,\,\,,
\end{equation}
where $\nu=1-du_+$. For a value of $\omega_{\rm max}$ found experimentally, 
this relation enables one to express the relaxation time $\tau_+$ by $u_+$ 
and the cell thickness $d$. Next, introducing the rates 
$\eta(\omega)=\varepsilon'(\omega)/\varepsilon''(\omega)$ and 
$\sigma(\omega)=\omega/\omega_{\rm max}$, and applying (34) and (35), leads to 
the following biquadratic equation with respect to $\nu$:
\begin{equation}
\sigma^2p^2-\eta\sigma(1+\sigma^2)p+2(1-\sigma^2)=0
\end{equation}
with
\begin{equation}
p=\nu+\frac{1}{\nu}\,\,\,.
\end{equation}
As a result, solving Eq. (39) for different values of $\omega$ and for 
respective, experimentally obtained values of $\eta$, yields $u_+$ as a 
function of $\omega$. Since $u_+$ is essentially independent of $\omega$, 
this function should be constant or should at least exhibit a plateau in some 
range of frequency. Thus, the considered procedure enables one not only to 
determine the surface energy parameter $u_+$, but also to test the above 
theoretical description of the dielectric response of chevron systems. In 
particular, one can determine the frequency region where this description is 
valid.   
\section{Experimental}
To verify the presented theory, dielectric measurements on the 
liquid-crystalline mixture Felix 15-100, commercially available from 
Clariant have been carried out. The investigated mixture exhibits 
ferroelectric smectic ${\rm C^*}$ phase at room temperature and transforms 
into the paraelectric smectic A phase at $73$ $^{\rm o}{\rm C}$. The measurements 
presented in this paper were performed at temperature $50$ $^{\rm o}{\rm C}$, 
i.e., $23$ $^{\rm o}{\rm C}$ below the ferroelectric-paraelectric phase 
transition. The investigated material was introduced into commercially 
available measuring cells of various thicknesses ($5$ $\mu{\rm m}$, from Linkam, 
UK, and $12.1$ $\mu{\rm m}$, $25.9$ $\mu{\rm m}$, $50$ $\mu{\rm m}$, from EHC, 
Japan). The thickness of all cells used was less than or comparable to the 
helical pitch of the investigated material ($\sim 20$ $\mu{\rm m}$), thus the 
chevron texture was easily attained. The walls of measuring cells were 
provided with indium-tin-oxide (ITO) semitransparent electrodes coated with 
thin polymer layers. These layers were rubbed unidirectionally along 
antiparallel directions on both plates. This promoted the orientation of 
molecules more or less parallel to electrodes and reduced the ionic current 
flowing across the sample. The measuring cell was placed inside a modified 
Metler hot stage, in which the temperature was controlled using a Digi-Sense 
temperature controller. The hot stage was installed between crossed polarizers 
of a polarizing microscope (Biolar, from PZO, Poland). The angle between the 
polarization plane of the incident light and the optic axis of the sample in 
the smectic A phase was set to $22.5^{\rm ^o}$. The intensity of light passing 
the sample and polarizers was registered using a photodiode connected to a 
preamplifier and lock-in amplifier SR 850 from Stanford Research (USA). 
Simultaneously, the electric capacitance of the sample was measured using the 
Hewlett-Packard low-frequency impedance analyzer HP 4192A. The measurements 
indicated that the dielectric increment (the difference between the electric 
permittivity in the smectic ${\rm C^*}$ phase and in the smectic A phase) was 
exactly proportional to the light modulation depth, detected by the photodiode, 
as it was already demonstrated in Ref.\cite{kc}. Thus, the dielectric 
measurements might be well replaced by electro-optic measurements, which are 
more accurate. Moreover, they are less sensitive to ionic current, especially 
at low frequencies\cite{kc}. This property enabled one to extend the measurement 
range down to $\sim 10$ Hz. In this way, reliable results were obtained at much 
lower frequencies than it might be possible using typical dielectric methods. 
Therefore, in what follows, electro-optic experimental results obtained for 
frequencies greater than 10 Hz are only discussed.
\section{Results and discussion}
The approach developed here to analyze the dielectric response of SSFLC with 
chevron geometry requires determining the effective surface interaction $u_+$. 
Using electro-optic data obtained for samples of different thicknesses and 
employing the procedure of Sec. II C, the parameter $u_+$ has been derived as a 
function of frequency in cases of various values of the voltage amplitude $U$. 
Plots of $u_+$ are shown in Fig. \ref{f2} for a low voltage, $U=0.1$ ${\rm V}$. 
It is seen that, for samples of different thicknesses, $u_+$ remains nearly 
constant within rather wide frequency ranges, and quickly decreases outside these 
ranges. Clearly, the rapid decay of $u_+$ in the low-frequency region is a 
consequence of an overlapping, in the dielectric response, between 
contributions arising from collective azimuthal excitations of molecules in 
chevron slabs and contributions that have other origins than these collective 
excitations. Similarly, a less sudden downfall of $u_+$ in high-frequency 
regions can be ascribed to components of $\varepsilon(\omega)$ corresponding 
to high-frequency processes\cite{go}, separated to a large extent from the 
collective processes occurring at intermediate frequencies. Naturally, the 
region of frequencies for which the parameter $u_+$ remains approximately 
constant determines the range of the validity of the introduced theoretical 
approach.

The surface energy parameter $u_+$ has been found for cells of different 
thicknesses by averaging parameter values calculated for frequencies 
belonging to ranges in which these values are nearly constant. Results of 
calculations of $u_+$ for $U=0.1$ ${\rm V}$ are given in Table I, where 
corresponding values of the ratio $\nu$ and the frequency 
$f_{\rm max}=\omega_{\rm max}/2{\rm \pi}$ are also given. Values of $u_+$ 
determined for various voltages demonstrate that $u_+$ scarcely depends on the 
voltage amplitude if $U$ remains relatively small, below some value 
$U_{\rm lin}$, different, in general, for individual samples 
($U_{\rm lin}\approx 3$ ${\rm V}$ for studied samples of the thickness 
$d=5$ $\mu{\rm m}$). When the voltage amplitude exceeds this value, nonlinear 
effects in $\varepsilon(\omega)$ become significant and Eq. (39) loses real, 
positive solutions for medium frequencies. As is seen from Table I, values of 
$\nu$, obtained in the low-voltage regime, are much greater than the threshold 
value, above which $\varepsilon''(\omega)$ has a single maximum. This means 
that, for the investigated SSFLC samples, differences between anchoring 
energies on both boundary plates do not strongly affect the form of the 
function $\varepsilon''(\omega)$. The dielectric response of studied chevron 
cells displays a dependence on the sample thickness. In particular, 
$f_{\rm max}$ decreases and thereby both the relaxation times $\tau_\pm$ 
increase, as $d$ grows. The increase of $\tau_\pm$ with $d$ is a consequence 
of a stabilization of collective reorientations of molecules by strong surface 
interactions. Since these interactions force the molecular ordering in entire 
chevron cells, time needed to reach the ordered state in initially perturbed 
samples is longer and longer as $d$ grows and/or as the surface interactions 
become weaker. This is consistent with relation (37). Clearly, the 
dependence of the effective surface interactions $u_\pm$ on the sample 
thickness (as seen in Fig. 2 in the case of $u_+$) originates, in general, 
from the dependence of surface energy strengths $\gamma_1$ and/or $\gamma_2$ 
on $d$. However, to explain such a behavior of $\gamma_1$ or $\gamma_2$ as $d$ 
is varied, one would analyze dynamic processes leading to the appearance of 
surface interactions.   
 
To test the method introduced here for investigating dielectric response of 
chevron cells, dielectric spectra have been determined using Eqs. (30), (34), 
(35), (37), and (38) with values $\omega_{\rm max}$ and 
$\varepsilon''(\omega_{\rm max})$ found experimentally. Additionally, a simple 
relaxation model that involves a single Debye process, characterized by the 
same values of $\omega_{\rm max}$ and $\varepsilon''(\omega_{\rm max})$, has 
been considered. Both resulting theoretical dielectric loss spectra as well as 
experimental spectra obtained for the voltage amplitude $U=0.1$ ${\rm V}$ are 
plotted in Fig. \ref{f3} for intermediate frequencies. These plots show that 
the experimentally determined dielectric loss spectra exhibit a broadening 
near their maxima, compared to corresponding spectra obtained for single Debye 
relaxation processes, and that such a broadening is well described by the 
introduced model including two Debye processes. Consequently, the effect of 
smearing of $\varepsilon''(\omega)$ near its maximum, located in the 
intermediate frequency region, can be considered as a result of a difference 
between anchoring energies at boundary surfaces. However, when $\omega$ 
decreases starting from $\omega_{\rm max}$, there appears a discrepancy 
between measured and theoretically predicted dielectric loss permittivities, 
as illustrated in Fig. \ref{f4}. This inconsistency is due to the occurrence 
in chevron samples of molecular motions other than the collective molecular 
reorientations taken into account in the theoretical model. Such additional 
motions are called here low-frequency processes, although they markedly affect 
the spectra $\varepsilon(\omega)$ even for $\omega$ relatively close 
to $\omega_{\rm max}$. Thus, frequency regions, in which low- and 
medium-frequency processes are significant, are not quite separated, and a 
rigid determination of the corresponding low- and intermediate-frequency 
ranges is not possible.

To interpret the complicated low-frequency behavior of the dielectric loss 
permittivity, it has been suggested that various dynamic mechanisms can be of 
importance in low-frequency regime\cite{p2,hv}. However, no detailed analysis 
of the complex dielectric response of chevron cells at low frequencies has 
been reported yet. For low voltages (for which the dielectric response is 
linear) contributions to $\varepsilon(\omega)$ that arise from dynamic 
processes other than fluctuations of the azimuthal angle with chevron slabs 
can easily be investigated by a subtraction of theoretical functions 
$\varepsilon_t(\omega)$ involving two relaxation times from respective 
experimental spectra $\varepsilon_e(\omega)$. Resulting extracted loss spectra 
$\Delta\varepsilon''=\varepsilon_e''-\varepsilon_t''$ are shown in 
Fig. \ref{f5} for samples of different thicknesses, in the case of 
$U=0.1$ ${\rm V}$. It proves that, for each system under study here, 
$\Delta\varepsilon''$ has a single maximum at a frequency $\omega_0$ of the 
same order of magnitude. Evidently, $\Delta\varepsilon''$ displays for 
$\omega>\omega_0$ a complex form, different from the shape of loss spectra 
corresponding to single Debye processes or even to sums of a few single 
processes. This suggests that the dielectric response of chevron cells can 
approximately be described by a spectrum of Debye processes with a continuous 
distribution of relaxation times. Then, for low frequencies, such that 
$\omega\ge\omega_0$, the extracted loss permittivity can be expressed as 
\begin{equation}
\Delta\varepsilon''(\omega)=\frac{1}{\omega}
\int^{\tau_2}_{\tau_1}\!\!\rho(\tau)\frac{\omega\tau}{1+\omega^2\tau^2}d\tau\,\,\,,
\end{equation}
where $\tau_2\le 1/\omega_0$ and $\tau_1$ are respectively maximal and minimal 
relaxation times, and $\rho(\tau)$ denotes the distribution of relaxation times. 
For $\omega\ge\omega_0$, the shape of the extracted spectra determined with the 
use of experimental data can easily be recovered by assuming that
\begin{equation}
\rho(\tau)\sim\tau^{-2}\,\,\,.
\end{equation}
Thus, applying (41) and (42) yields
\begin{equation}
\Delta\varepsilon''(\omega)=C_1\ln{\Big[}\frac{\tau_2^2(1+\omega^2\tau_1^2)}
{\tau_1^2(1+\omega^2\tau_2^2)}{\Big]}
\end{equation}
with $C_1$ being a positive constant. Similarly, using (42), one finds the 
extracted dielectric permittivity $\Delta\varepsilon'$ to be
\begin{eqnarray}
\Delta\varepsilon'(\omega)=C_2{\big[}&&\frac{1}{\omega}{\big(}\frac{1}{\tau_1}
-\frac{1}{\tau_2}{\big)}\nonumber\\
&&-\arctan(\omega\tau_2)+\arctan(\omega\tau_1){\big]}\,\,\,,
\end{eqnarray}
where $C_2$ denotes a constant. Results of fits of functions given by Eqs. (43) 
and (44) (on adjusting the parameters $C_1$, $C_2$, $\tau_1$, and $\tau_2$) to 
respective extracted spectra obtained with the use of experimental data are 
shown in Figs. \ref{f5} and \ref{f6}. It is seen that the agreement between 
postulated and experimental extracted spectra $\Delta\varepsilon''$ is good 
(as long as $\omega\ge\omega_0$), but is less satisfactory in the case of the 
spectra $\Delta\varepsilon'$. Nevertheless, these figures demonstrate that the 
low-frequency dielectric response of chevron cells can indeed be characterized 
by Debye processes with continuously distributed relaxation times. It should 
also be noted that the measured spectra presented in Fig. \ref{f5}, contrary 
to earlier results derived from a numerical decomposition of the dielectric 
loss spectra\cite{ka}, do not reveal the existence of any contributions that 
would correspond to surface fluctuations of the $c$ director (at frequencies 
of an order magnitude greater than $\omega_{\rm max}$).       

Turning to observations of chevron samples through polarizing optical 
microscope, it is argued here that a complex dielectric response of SSFLC is 
connected to dynamics of zigzag defects\cite{la}. Usually, these defects 
appear spontaneously, forming irregular patterns, when chevron cells are 
cooled down\cite{la,h2,li,fu,wa}. Microscopic observations of chevron samples 
performed in the low-frequency regime at weak voltages have displayed periodic 
changes (according to a voltage alternation) of the color and the intensity of 
light transmitted not only by uniform chevron regions (domains) but also by 
thin and thick zigzag walls separating these regions, as illustrated in the 
microphographs of Fig. \ref{f7} for a sample of the thickness 
$d=5$ $\mu{\rm m}$. Thus, in addition to uniform domains, collective relaxation 
processes caused by a weak voltage alternating with low frequencies occur also 
inside zigzag walls of different widths. Since zigzag walls reveal various 
shapes as well as sizes\cite{la,c2,li} and consist of smectic A layers having 
different thicknesses, Consequently, being affected by surface 
interactions with different strengths, one can infer that relaxation times 
characterizing dynamic processes within these walls undergo a nonuniform 
distribution. Then, the complexity of the dielectric response of chevron cells 
to weak voltages alternating with low frequencies may, in fact, be considered as 
a result of relaxation processes within zigzag walls. 

As the amplitude of the applied voltage exceeds a threshold $U_1$ (different, 
in general, for various samples), thick zigzag walls start to creep. Such a 
drift viscous motion of defect walls, induced by an alternating electric 
field, is spatially confined by pinning ends of thick walls to ends of thin 
walls. This is illustrated in Fig. \ref{f8} for the same sample of the width 
$d=5$ $\mu{\rm m}$, in the case of the voltage amplitude $U=1.0$ ${\rm V}$ and 
the frequency $f=100$ ${\rm Hz}$. The drift motion of thick walls 
[Figs. \ref{f8}(a) and \ref{f8}(b)] is not periodic, although long time 
averages of displacements $r(t)$ of these walls vanish, i.e., 
$\lim_{t\rightarrow\infty}r(t)=0$, as shown in Fig. \ref{f8}(c). 

When the voltage amplitude goes beyond an another threshold $U_2$ ($U_2>U_1$), 
both thin and thick walls begin to slide. Such irreversible motions of zigzag 
walls are accompanied with a gradual shrinkage of corresponding zigzag lines 
on boundary surfaces and with gradual disappearing of the defect walls, as $U$ 
grows, remaining below the switching threshold $U_s$. Thus, the density of 
zigzag defects decreases when $U$ grows (above the sliding threshold $U_2$). 
The effect of motions of zigzag walls in the sliding regime is presented in 
Fig. \ref{f9} for the studied sample of the thickness $d=5$ $\mu{\rm m}$, in 
the case of the voltage amplitude $U=5$ ${\rm V}$ and the frequency 
$f=100$ ${\rm Hz}$. Note that the threshold voltage at which the switching 
process appears, found for samples of different thicknesses, increases with 
$d$ (e.g., $U_s=10$ ${\rm V}$ for the case of $d=5$ $\mu{\rm m}$, while 
$U_s=20$ ${\rm V}$ for the case of $d=25.9$ $\mu{\rm m}$). Analysis of the 
dynamics of defects presented here indicates that both the threshold voltages 
$U_1$ and $U_2$ (similarly to $U_s$) are different for samples of different 
thicknesses. However, to answer the question of whether the dependence of these 
threshold voltages on the sample thickness has a general character, and to 
possibly determine such a general dependence, further work on a wider class 
of uniform chevron cells is needed.       

Changes in the dynamic behavior of defect walls as $U$ crosses its threshold 
values are markedly reflected in the low-frequency dielectric loss spectra, 
as illustrated in Fig. \ref{f10}. It is seen that, in comparison to 
relaxation processes inside zigzag walls, creep motions of thick walls give a 
large contribution to $\varepsilon''(\omega)$, in almost the entire 
low-frequency range, with a distinct maximum. Contrary to creep loss spectra, 
sliding spectra (associated with sliding of defect walls) have very small 
amplitudes for $f>10$ ${\rm Hz}$. This is a consequence of the fact that 
sliding motions are very slow compared to creep motions. Hence, the 
dielectric loss is dominated in the sliding regime by the low frequency tail 
of contributions to $\varepsilon''(\omega)$ arising from collective 
fluctuations of the azimuthal orientation of the $c$ director. It should be 
noted that similar relaxation, creep, and slide motions of complex objects in 
elastic media have recently been studied, both theoretically and experimentally, 
in various contexts. In particular, thermally activated and/or field-induced 
domain wall motions of these types have been investigated in impure 
magnets\cite{nt}, polydomain relaxor-ferroelectric single crystals\cite{kl}, 
discontinuous metal-insulator multilayer systems\cite{ch}, and periodically 
poled ferroelectric single crystals\cite{ba}. Transitions between different 
dynamic states of domain walls occurring in these systems have been shown to 
reveal a character of dynamic phase transitions, at critical field amplitudes 
being some functions of the temperature $T$ and the frequency of the field 
alternation. Clearly, one can expect that changes in the dynamics of zigzag 
walls appearing in chevron SSFLC also display a character of dynamic phase 
transitions, determined in the parameter space ($U-T-\omega$). A detailed 
exploration of this question is, however, beyond the scope of this paper.     
\section{Conclusions}
It has been shown in this paper that the dielectric response of ferroelectric 
chevron systems stabilized by nonpolar and polar surface interactions is 
determined at intermediate frequencies by two Debye relaxation processes, 
corresponding to collective fluctuations of the azimuthal orientation of 
molecules in two chevron slabs. This relaxation mechanism has been proved to 
cause a broadening of the dielectric loss spectrum near its maximum, compared 
to a spectrum obtained for a model involving a single Debye process. Such a 
broadening effect has been confirmed experimentally, for chevron cells of 
different thicknesses. To investigate low-frequency processes and possible 
intermediate-frequency processes, other than cooperative fluctuations of 
azimuthal orientations of the $c$ director, theoretically derived dielectric 
response functions have been extracted from respective measured electro-optic 
spectra. Resulting extracted spectra have not displayed any additional 
processes at medium frequencies, even for relatively thick samples. This is 
in contrast to an earlier suggestion\cite{ka} that, in a medium-frequency 
region, the dielectric response of thick enough cells can be affected by 
surfacelike fluctuations of the $c$ director. It has been demonstrated that 
low-frequency dynamic processes have different character in three voltage 
regimes and can be ascribed to relaxation processes inside thin and thick 
zigzag walls, creeping of thick walls, and sliding as well as disappearing 
of defect walls. Certainly, low-frequency dielectric response of chevron 
cells can be affected by other kinds of defects or smectic layer 
deformations\cite{la,fu,is}. However, the low-frequency dynamic mechanisms 
discussed in this paper appear to be most important. Finally, it should be 
remarked that the possibility of macroscopic observations of dynamic 
behaviors of defect walls provides a useful tool of studying low-frequency 
dynamic states of chevron SSFLC and dynamic phase transitions between these 
states.  
\acknowledgments 
The authors are very grateful to J. Ma{\l}ecki for useful discussions. 
This work was supported by Polish Research Committee (KBN) under grant 
No. 2PO3B 127 22.

%
\begin{figure}
\caption{Geometry of orientation states in a uniform chevron cell of thickness 
$d$: (a) in the smectic layer plane ($X$-$Y$), and (b) in the ($X$-$Z$) plane 
(perpendicular to the smectic layer plane and to the bounding plates). In (a) 
the symbols $\rightarrow$ and $\vdash$ represent, respectively, projections of 
the microscopic polarization $\vec{P}_s$ and the $c$ director $\vec{c}$ on the 
plane ($X$-$Y$). In (b) the symbols $\rightarrow$ and $>\!\!\!-$ denote 
respectively projections of $\vec{P_s}$ and the molecular director $\vec{n}$ 
on the plane ($X$-$Z$). Scales of projections of vectors $\vec{c}$ and $\vec{n}$ 
on the planes ($X$-$Y$) and ($X$-$Z$), respectively, are not preserved. The 
abbreviations $T$ and $B$ indicate top and bottom bounding surfaces, 
respectively, whereas $I$ denotes the interface plane. The azimuthal angle 
$\phi$ takes pretilt values $\phi=\pm\phi_0$ in upper and lower chevron slab, 
respectively.}
\label{f1}
\end{figure}
\begin{figure}
\caption{The surface interaction parameter $u_+$ determined as function of 
the frequency of the alternating voltage of the amplitude $U=0.1$ ${\rm V}$, for 
samples of different thicknesses: $d=5$ ${\rm\mu m}$ $(\bigcirc)$, 
$d=12.1$ ${\rm \mu m}$ $(\triangle)$, $d=25.9$ ${\rm \mu m}$ $(\Box)$.}
\label{f2}
\end{figure}
\begin{figure}
\caption{Loss spectra determined experimentally (dots) within a medium 
frequency range, for $U=0.1$ $V$, in cases of chevron cells of thicknesses 
$d=5$ ${\rm\mu m}$ and $d=12.1$ ${\rm\mu m}$, and derived theoretically by the 
use of the approach including two Debye processes (labeled 1) as well as by the 
use of the single relaxation model (labeled 2). Theoretical plots were obtained for 
respective, measured values of 
$\omega_{\rm max}$ and $\varepsilon''(\omega_{\rm max})$.}
\label{f3}
\end{figure}
\begin{figure}
\caption{Experimental (dots) and theoretical (solid lines) loss spectra for 
chevron cells of different thicknesses. Theoretical plots were obtained by the 
use of the approach involving two Debye processes.}
\label{f4}
\end{figure}
\begin{figure}
\caption{Dielectric loss spectra obtained by a subtraction of theoretical loss 
spectra $\varepsilon_t$ from experimental spectra $\varepsilon_e$, in cases of 
samples of different thicknesses, for $U=0.1$ ${\rm V}$. Dots correspond to 
experimental data, while the solid line refers to the postulated function of 
Eq. (43).}
\label{f5}
\end{figure}
\begin{figure}
\caption{Extracted spectra $\Delta\varepsilon'$ obtained for a sample of the 
thickness $d=5$ $\mu{\rm m}$, in the case of $U=0.1$ ${\rm V}$. Dots refer to 
experimental data and the solid line represents the function given by 
Eq. (44).}
\label{f6}
\end{figure}
\begin{figure}
\caption{(Color online) Zigzag walls in a chevron cell of the thickness 
$d=5$ $\mu{\rm m}$, under the influence of a voltage of the amplitude 
$U=0.5$ ${\rm V}$, alternating with the frequency $f=10$ ${\rm Hz}$. The 
microphotographs were taken at a short time period, both at the speed shutter 
$1/100$ ${\rm s}$. These micrographs give an evidence of a pulsation of light 
transmitted not only by domains but also by defect walls of various widths.}
\label{f7}
\end{figure}
\begin{figure}
\caption{(Color online) Creep motions of thick zigzag walls in a sample of 
the thickness $d=5$ $\mu{\rm m}$, induced by an alternating voltage of the 
amplitude $U=1$ ${\rm V}$ and the frequency $f=100$ ${\rm Hz}$. The 
microphotographs (a) and (b) were taken at the time period $1$ ${\rm s}$, both 
at a high shutter speed $1/1000$ ${\rm s}$. Thin walls, seen on left and 
right sides of these micrographs, remain fixed confining the motion of the 
thick wall connected to the thin walls. (c) is a micrograph of the same 
objects as in micrographs (a) and (b) but taken at a slow shutter speed 
$1$ ${\rm s}$. This figure shows a thick-wall drift, averaged over a relatively 
long time period.} 
\label{f8}
\end{figure}
\begin{figure}
\caption{(Color online) Sliding of zigzag walls in a chevron sample of the 
thickness $d=5$ $\mu{\rm m}$, for the voltage amplitude $U=5$ ${\rm V}<U_s$ and 
the frequency $f=100$ ${\rm Hz}$. The micrographs (a) and (b) were taken at the 
shutter speed $1/2000$ ${\rm s}$, in a time $1$ ${\rm s}$ [(b) after (a)]. The 
arrows indicate an irreversible motion and a shortening of respective defect 
walls in this lapse of time.}
\label{f9}
\end{figure}
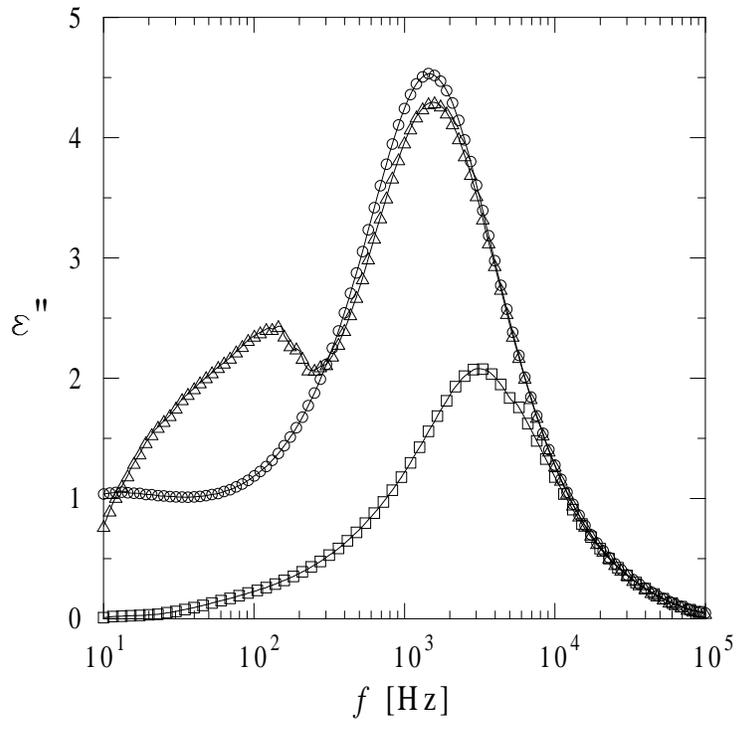
\begin{figure}
\caption{Experimental loss spectra obtained for a chevron system of the 
thickness $d=5$ $\mu{\rm m}$, in cases: $U=0.3$ ${\rm V}<U_1$ ($\bigcirc$), 
$U_1<U=1.0$ ${\rm V}<U_2$ ($\triangle$), and $U_2<U=5$ ${\rm V}<U_s$ ($\Box$). 
The shift of the maximum of $\varepsilon''$ in the case of $U=5$ ${\rm V}$ 
gives evidence of nonlinear effects in collective azimuthal excitations 
of molecules.}
\label{f10}
\end{figure}
\begin{table}
\caption{Values of $u_+$, $\nu$, and $f_{\rm max}$ determined for chevron 
samples of different thicknesses. The parameters $u_+$ and $\nu$ were 
obtained by applying the procedure described in Sec. II and by using 
electro-optic data.
\label{t1}}
\begin{tabular}{cccccccccc}
&$d\,\,[{\rm\mu m}]$& \hspace{14mm} &$u_+\,\,[{\rm m^{-1}}]$& \hspace{7mm} 
& $\nu$& \hspace{7mm} & $f_{\rm max}\,\,[{\rm Hz}]$& \vspace{1mm}\\
\tableline
\vspace{-3mm}\\
&\,\,\,$5.0\,$& &$6.95\times 10^4$& &$0.348$& &$1349.3$&\\
&$12.1\,$& &$3.67\times 10^4$& &$0.556$& &$\,\,\,363.1$&\\
&$25.9\,$& &$1.89\times 10^4$& &$0.510$& &$\,\,\,186.5$&\\
&$50.0\,$& &$1.21\times 10^4$& &$0.395$& &$\,\,\,208.9$&
\end{tabular}
\end{table}
\end{document}